\pdfoutput=1

\documentclass[preprint,5p,twocolumn,times]{elsarticle}

\UseRawInputEncoding
\usepackage{graphicx}
\usepackage{subfigure}
\usepackage{url}
\usepackage{amssymb}
\usepackage{amsmath}
\usepackage{amsfonts}
\usepackage{wrapfig}
\usepackage{afterpage}
\usepackage{float}
\usepackage{color}
\usepackage{epstopdf}
\usepackage{bm}
\usepackage{leftidx}
\usepackage{hyperref}
\usepackage{array}
\usepackage{multicol}
\usepackage{multirow}
\usepackage{booktabs}
\usepackage[figuresright]{rotating}
\usepackage[linesnumbered, ruled,vlined]{algorithm2e}
\usepackage[noend]{algpseudocode}

\graphicspath{{doc_figs/}}

\newcommand{\brho}{\mbox{\boldmath$\rho$}}

\newcommand{\bsig}{\mbox{\boldmath$\sigma$}}
\newcommand{\bsigma}{\mbox{\boldmath$\sigma$}}

\newcommand{\bphi}{\mbox{\boldmath$\phi$}}

\newcommand{\bvarepsilon}{\mbox{\boldmath$\varepsilon$}}
\newcommand{\bchi}{\mbox{\boldmath$\chi$}}

\newcommand{\bveps}{\mbox{\boldmath$\varepsilon$}}

\def\cD{\mathbf{D}}

\def\a{\mathbf{a}}

\def\b{\mathbf{b}}

\def\e{{\bf e}}
\def\f{\mathbf{f}}
\def\bff{\mathbf{f}}
\def\g{\mathbf{g}}

\def\bk{\mathbf{k}}

\def\p{\mathbf{p}}
\def\bp{\mathbf{p}}

\def\bu{\mathbf{u}}

\def\bv{\mathbf{v}}
\def\v{\mathbf{v}}
\def\x{\mathbf{x}}

\def\B{{\bf B}}
\def\C{\mathbf{C}}
\def\D{\mathbf{D}}
\def\bD{\mathbf{D}}

\def\E{\mathbf{E}}

\def\F{\mathbf{F}}

\def\G{\mathbf{G}}

\def\bI{\mathbf{I}}
\def\I{\mathbf{I}}
\def\J{\mathbf{J}}
\def\bJ{\mathbf{J}}
\def\bK{\mathbf{K}}

\def\K{\mathbf{K}}
\def\N{\mathbf{N}}

\def\P{\mathbf{P}}
\def\bP{\mathbf{P}}

\def\Tr{\mathrm{Tr}}
\def\rmT{\mathrm{T}}

\def\RR{\mathbb{R}}

\newcommand{\real}{\mathbb{R}}

\def\calU{{\cal U}}

\def\calM{{\cal M}}
\newcommand{\bmu}{\mbox{\boldmath$\mu$}}

\def\eb{\begin{equation}}
\def\ee{\end{equation}}

\begin{document}

\begin{frontmatter}
\title{Lattice structure design optimization under localized linear buckling constraints } 

\author[cadaddress]{Xingtong Yang$^{1,}$}
\author[cadaddress]{Xinzhuo Hu$^{2,}$}
\author[cadaddress]{Liangchao Zhu$^{3,}$}
\author[cadaddress]{Ming Li\corref{mycorrespondingauthor}}
\cortext[mycorrespondingauthor]{Corresponding author: liming@cad.zju.edu.cn}
\address[cadaddress]{State Key Laboratory of CAD\&CG, Zhejiang University, Hangzhou, China}

\begin{abstract}
An optimization method for the design of multi-lattice structures satisfying local buckling constraints is proposed in this paper. First, the concept of free material optimization is introduced to find an optimal elastic tensor distribution among all feasible elastic continua. By approximating the elastic tensor under the buckling-containing constraint, a matching lattice structure is embedded in each macro element. The stresses in local cells are especially introduced to obtain a better structure. Finally, the present method obtains a lattice structure with excellent overall stiffness and local buckling resistance, which enhances the structural mechanical properties.

\end{abstract}
\end{frontmatter}

\section{Introduction}
Lattice structures have fascinating properties of lightweight and multi-functions such as shock resistance~\cite{elnasri2007shock}, energy absorbability~\cite{2014Designing}, damping enhancement~\cite{golovin2003damping} and defect tolerance~\cite{andrews2001influence}. These prominent properties make them an ideal candidate in widespread industrial applications including aerospace, automotive, and biomedical field~\cite{2011Additive,2015Additive}. Recent developments and applications of the lattice structures are further referred to the reviews from Dong et al.~\cite{Dong2018A} and Tamburrino et al.~\cite{2018LatticeRev}.

Automatic design of a complex lattice structure to meet critical mechanical requirements on the other hand becomes a dominating task. Topology optimization is becoming a powerful automatic design tool that computes the optimal material distribution to meet certain design target and constraints, and various excellent approaches have been proposed over the past decades~\cite{Sigmund1994Materials,Sigmund1995Tailoring,1999Optimization}. The approaches mainly depend a homogenization-based framework, where topologies of the micro-structures or their distributions are optimized for overall structural performance optimization, or for the identification of periodic microstructure shapes attaining the theoretical limit on elastic properties~\cite{Sigmund1997Design,Gibiansky2000Multiphase,Sigmund2000A,Neves2000Optimal,Sigmund1999On}.

However, an extreme stiff lattice cell or structure may still fail as the appearance of slender lattice bars are very vulnerable to highly localized buckling when being imposed by compressive loading. Actually, it has been found that the strength design achieved in industrials is still far from the theoretical strength limit it can achieve~\cite{Bauer2016Approaching}. Noticing that the buckling design often has a counter-acting character to compliance or stress designs, it must be carefully incorporated into the optimization of such complex structures.

Topology optimization of a lattice structure under buckling constraint is however a more challenging topic than solely considering stiffness optimization. The challenges mainly come from its much more complex computational efforts, and the poor convergence. The computational costs are due to its much more complex eigenvalue analysis problem especially for large-scale analysis problem. The convergence issue comes from the typical challenges encountered in the eigenvalue optimization problem, including spurious local buckling modes~\cite{Neves1995Generalized,neves2002topology}, or the eigenvalues with multiplicity .

Previous work on topology optimization under buckling constraint mainly focus on macro-scale optimization or periodic microstructure optimization. Topology optimization of continuum structures with a global buckling criterion has been long considered by Neves et al~\cite{Neves1995Generalized}, and later including geometrically non-linear response in~\cite{kemmler1999topology,Buhl2000Stiffness,Bruns2001Topology}.

On the other hand, the problem of designing a microstructure under buckling constraint was initially considered by Bendsoe and Triantafyllidis [32] as a size optimization problem for a rectangular orthogrid. Subsequently, a topology optimization formulation for microstructure design under buckling constraint was formulated by Neves et al.~\cite{Neves1995Generalized}, which only treated localized instability modes. The study was later extended to cover non-local modes with the Bloch-wave technique in~\cite{Neves2001Analysis,neves2002topology}.
Recently, Thomsen et al.~\cite{Sigmund2018Buckling} considers the design of periodic microstructures with respect to multi-scale buckling conditions under typical different stress situations, laying the foundations for future multi-scale structural and material design. Some previous research efforts have also been devoted to design microstructures in balancing the conflicts between the stiffness and stability~\cite{Gao2020Improving,2020Extreme,2020Numerical}.

In spite of these previous studies, designing a lattice structure to maximally resist the local buckling while simultaneously maintaining the global structure stiffness seems to be even empty as far as we know, which are to be explored in this study. The optimization framework has to resolve critical issues as how to resist local buckling without much losing the global structural stiffness. This is achieved in a two-state process by first computing optimal material distribution and then embedding microstructures under buckling constraints. Specifically, we first compute a optimal distribution of the material elasticity tensor associated to each macro-element based on an approach of free (isotropic) material optimization. The free material properties are further clustered into a small group of different elasticity tensors based on a machine learning technique to accelerate downstream tasks. After this, the lattice structure is to be found for each macro-element by approximating the elasticity tensor under the buckling constraint. Here, the location stress is also included here for an further improvement of performance optimization based on an observation that the buckling mode does not only depend on its shape of microstructures but also on its external loadings.

\section{Optimal material property field generation in macroscale}

Given a design domain under certain boundary conditions, the widely studied problem of minimum compliance, or equivalently maximum stiffness, is examined here. The problem of FMO
is first studied here to provide an optimized field of material property. In order for an appropriate and manufacturable interpretation for follow-up microstructure design, manufacturability constraint is further introduced in the FMO problem following previous study.
Here the \emph{manufacturability} refers to an elasticity tensor which can be obtained as an effective material property of a solid-void microstructure.
After this, a novel material clustering approach is also developed that greatly reduces the material space while maintaining high compliance fidelity of the generated structures. Details are explained below.

\subsection{Obtain optimal material distribution}
The problem of FMO takes the material elasticity tensor in each domain point as design variables, and finds their optimal distribution within the design domain. The 2D case is studied for ease of explanation. Consider a discrete macro- design domain $\Omega=\{\Omega_{e}, e=1,\ldots,M\}\subset \real^2$ made of disjoint square elements $\Omega_e$ of the same size. For each $\Omega_e$, let $\bsig_e$ and $\bveps_e$ be the second-tensor stress and strain vectors using the conventional Kelvins notation
{\small
\begin{align}
\bveps_e=(\bveps_{11},\bveps_{22},\sqrt{2}\bveps_{12})^T, \bsig_e=(\bsig_{11},\bsig_{22},\sqrt{2}\bsig_{12})^T,
\end{align}
}
where
{\small\begin{equation}
\bveps_e=\frac{1}{2}\left(\nabla \bu_e + \nabla \bu_e^T \right),
\end{equation}}
$\bu_e\in \RR^8$ is the associated displacement vector and $\nabla$ is the gradient operator.
The system studied here is governed by the Hooke's law, where the stress is a linear function of the strain as
{\small\begin{equation}
\bsig_e=\bD_e\bveps_e,
\end{equation}}
and the elasticity tensor $\bD_{e}$ is a symmetric positive $3\times3$ matrix as
{\small
\begin{align}
\bD_e =
\left[
\begin{array}{rrr}
\bD_{1111} & \bD_{1122} &\sqrt{2}\bD_{1112} \\
 & \bD_{2222} &\sqrt{2}\bD_{2212}\\
sym. &  & 2\bD_{1212}
\end{array}
\right].
\label{eq:Ci}
\end{align}
}

The free material optimization problem aims to find the optimal material property $\bD_e$ associated to each $\Omega_e$ such that the resulted structure is stiffest, that is,
find $\cD=(\bD_1, \ldots,\bD_M)$ so that the compliance of the resulted structure $(\Omega, \bD)$ is minimized, that is,
{\small
\begin{equation}\label{eq:FMO}
\underset{\cD_e\in\mathbb{S_+^N}}{\min} \ C(\cD,\bu)=\bff^T\bu ,\quad s.t.
\end{equation}
}
{\small
\[
\left\{
\begin{array}{lll}
&\bK(\cD)\bu=\bff,\ \bu\in \calU,  & \mbox{equilibrium equation} \\
&\sum_{e=1}^{M} \Tr(\bD_e)\leq T_0, & \mbox{global trace constraint} \\
&\underline{T} \leq \Tr(\bD_e)\leq \overline{T}, & \mbox{element trace constraints},
\end{array}
\right.
\]}
where $\bff$ is the exerted nodal force vector ignoring the structure weight for simplicity, and $\mathbb{S_+^N}$ is the cone of symmetric positive semidefinite matrices in the space $\mathbb{S^N}$ of symmetric $N \times N$ matrices for $N=3$ in 2D, $N=6$ in 3D. $\calU\subset \real^d$ is the admissible displacement space, where certain Dirichlet boundary conditions are prescribed. $\Tr(\bD_e)$, the trace of $\bD_e$, is used to denoted the cost of elemental material, which is bounded by $\underline{T}$ and $\overline{T}$. $T_0$ constrains the overall amount of materials distributed in the structure. The influence of the trace bound $\overline{T}$, $T_0$ on performance of the final design will be discussed in Section~\ref{sec-example}.

The global stiffness matrix $\bK(\cD)$ is calculated by
{\small\eb
\bK(\cD) = \sum_{e=1}^M \K_e(\cD),\ \K_e(\cD)=\sum_{k=1}^{n_G}\B_{e,k}^T \bD_e \B_{e,k},
\ee}
where $\B_{e,k}$ is the strain-displacement matrix and $n_G$ is the number of Gaussian integration points. Note that $\bK$ and $\bK_e$ are all symmetric positive semidefinite.
The objective function $c(\bD,\bu)$ is the structure's compliance calculated by
{\small\eb\label{eq-compliance}
C(\cD,\bu)
= \bu^{T} \bK(\cD) \bu
= \sum_{e=1}^M \bu_e^{T} \bK_e(\cD_e) \bu_e.
\ee}

The FMO problem~\eqref{eq:FMO} includes nonlinear and nonconvex vector constraints. Using the Schur complement theorem, this nonconvex semi-definite Programming (SDP) problem can be written as a linear SDP problem, which has a converged global optima~\cite{kovcvara2008free}
{\small
\begin{equation}\label{eq:Schur}
\underset{\bD_e\in\mathbb{S_+^N},\gamma>0 }{\min} \ \gamma ,\quad s.t.
\end{equation}}
{\small\[
\left\{
\begin{array}{ll}
&\begin{pmatrix}
    \gamma  & \bff^T\\
    \bff & \bK(\cD)
\end{pmatrix}\geq 0,\\
&\sum_{e=1}^{M} \Tr(\bD_e)\leq T_0,\\
& \underline{T} \leq \Tr(\bD_e)\leq \overline{T}, \  \ e=1,\ldots, M.
\end{array}
\right.
\]}

There is a notable fact that the original problem~\eqref{eq:Schur} will generate some extreme elasticity tensors which usually make the subsequent microstructure design difficult or even impossible. Therefore, additional manufacturability constraints are imposed here to bound the minimal eigenvalues of elasticity tensors away from zero, in case that any directional stiffness is too close to zero,
{\small\eb\label{eq:FMO_cons}
\bD_e - \delta\bI \geq 0, \ e=1,\ldots, M,
\ee}
where $\bI$ is the identity matrix. The value of the desired $\delta$ varies with different design problems, and can be computed using the method described in ~\cite{schury2012efficient}.

Solution to the above FMO problem can be computed efficiently via a primal-dual interior point method. Even though the solution to problem~\eqref{eq:Schur} becomes sub-optimal after adding constraint~\eqref{eq:FMO_cons}, the downstream task of microstructure embedding is facilitated~\cite{podesta2018material}.

In the application, some special requirements will be imposed on the properties of the material according to the actual situation. For example, the elastic tensor matrix of the orthotropic material needs to exceptionally satisfy the condition
\eb
\left[
\begin{array}{c}
    \sigma_x \\
    \sigma_y \\
    \tau_{xy}
\end{array}
\right]
=
\left[
\begin{array}{ccc}
    \frac{E_x}{1-\nu_{xy}\nu_{yx}} & \frac{\nu_{yx} E_x}{1-\nu_{xy}\nu_{yx}} & 0 \\
    \frac{\nu_{yx} E_y}{1-\nu_{xy}\nu_{yx}} & \frac{E_y}{1-\nu_{xy}\nu_{yx}} & 0 \\
    0 & 0 & G_{xy} \\
\end{array}
\right] \cdot
\left[
\begin{array}{c}
    \varepsilon_x \\
    \varepsilon_y \\
    \gamma_{xy}
\end{array}
\right]
\label{eq-ort}
\ee
and $ \nu_{xy} E_y  = \nu_{yx} E_x $.
Or
\eb
\C
=
\left[
\begin{array}{ccc}
    C_{1111} & C_{1122} & 0 \\
    C_{1122} & C_{2222} & 0 \\
    0 & 0 & C_{1212} \\
\end{array}
\right]
\label{eq-C-ort}
\ee
For isotropic material optimization problems, the methods described in our previous work can be consulted.

\subsection{Material space reduction via material clustering}

The large number of different kinds of materials generated by this process poses significant challenges for the practical application of the optimization results, such as the large computational costs associated with the inverse homogenization process, as well as the future manufacturing difficulties and the cost of testing the physical properties.
With the clustering method proposed in this section, the original excessive number of material types can be reduced to a specified number without excessive loss of overall physical properties of the resulting structures.

Hierarchical clustering groups data over a variety of scales by creating a cluster tree or dendrogram. The tree is not a single set of clusters, but rather a multilevel hierarchy, where clusters at one level are joined as clusters at the next level. In the proposed approach, the similarity between two elasticity tensors is measured by their Euclidean distance, and different design elements are iteratively grouped into a binary hierarchical cluster tree. The final K-clusters are derived via cutting the hierarchical tree into $K$ clusters $\Xi_k,\ k=1,\ldots,K$, and the corresponding cluster center is decided by performing the following FMO process, defined as,

{\small
\begin{equation}\label{eq:2nd-FMO}
\underset{\bD_e\in\mathbb{S_+^N},\gamma>0 }{\min} \ \gamma ,\quad s.t.
\end{equation}
\[
\left\{
\begin{array}{ll}
&\begin{pmatrix}
    \gamma  & \F^T\\
    \F & \bK(\bD)
\end{pmatrix}\geq 0,\\
&\sum_{k=1}^K \sum_{e\in \Xi_k} \Tr(\bD_e)\leq T_0,\\
&\underline{T} \leq \Tr(\bD_{k})\leq \overline{T}, \ k=1,\ldots, K\\
&\bD_{k} - \delta\bI \geq 0, \ k=1,\ldots, K,
\end{array}
\right.
\]
}
where the integer $K$ is the prescribed number of cluster groups. Note the trace sum of $\bD_i$ is still computed on all elements in the design domain.

\section{Lattice structure modeling in implicit form}\label{sec-latmodel}
Lattice structure modeling aims to build a global smooth lattice structure whose property is able to be analyzed via homogenization-based approach. It is specifically to address the following key challenges. Firstly, the lattice cell has a small number of control parameters to clearly describe its geometry. Secondly, the cells are varied in different elements but are globally and smoothly connected. Thirdly, the lattice cells have to cover as wide as possible a range of material space for performance improvement. Fourthly, each cell is able to be geometrically periodically distributed for the usage of numerical homogenization to predict its elasticity tensor. Details are explained below.

\begin{figure}[tb]
\centering
{\includegraphics[width=0.18\textwidth]{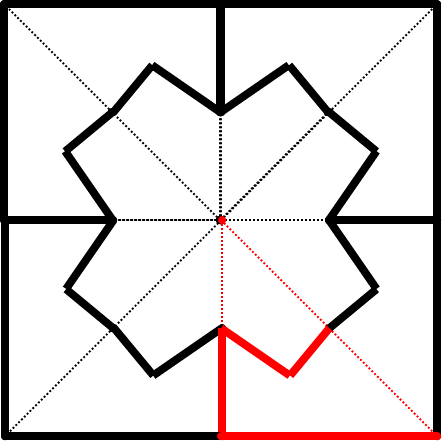}}\quad
{\includegraphics[width=0.2\textwidth]{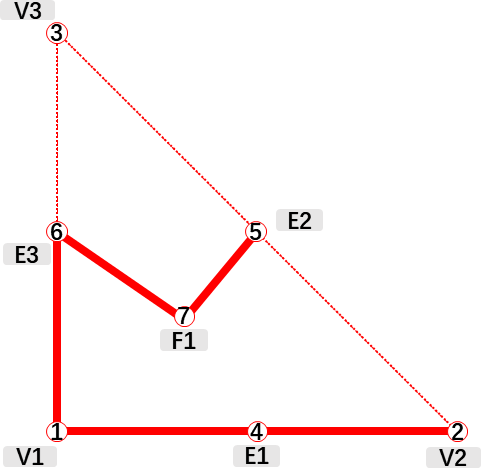}}
\caption{A schematic diagram of a complete microstructure model and its one-eighth.}
\label{pics:symmetry}
\end{figure}

\subsection{Lattice unit and structure modeling}
\begin{figure}[tb]
\centering
{\includegraphics[width=0.2\textwidth]{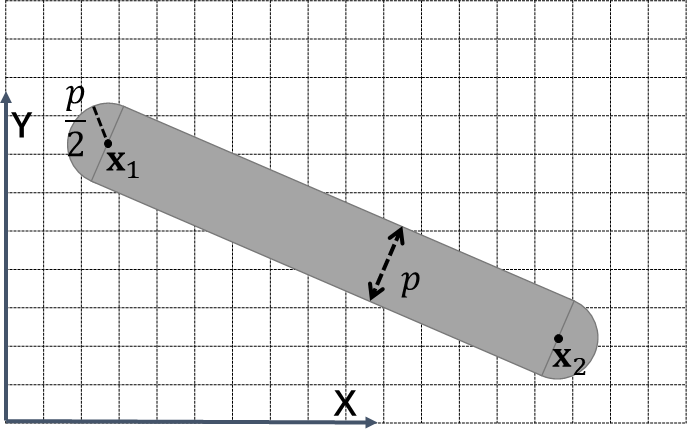}}
\caption{One bar determined by two nodes $\x_1,\x_2$ and the width $p$.}
\label{pics:single_bar}
\end{figure}

Basically, a lattice unit can be modeled in a parametric form, an implicit form, or a discrete mesh form. The implicit form has its merits in ease of freeform modeling, blending generation, boolean operation etc, and is taken here for lattice structure modeling. The lattice unit is defined by a set of bars of connecting nodes in a cubic element; its extension to more general rhombic elements is to be explained later.

The lattice unit is prescribed within a $1/8$ of a cubic element; a mirror symmetry gives the overall unit; see Fig.~\ref{pics:symmetry}. The following notations are used:
\\
$\Omega$: a given solid model; \\
$\calM$, a hexahedron mesh of $\Omega$; \\
$\bv^{(i)}_1,\bv^{(i)}_2$: coordinates of two endpoints of the bar $\iota^{(i)}$;\\
$p^{(i)}$: diameter of the bar $\iota^{(i)}$;\\
$\iota^{(i)}$: the $i$-th bar of the lattice;\\
$\omega$: the lattice unit consisting of $m$ bars; \\
$\Psi$: the overall lattice structure generated from $\calM$;\\
$\tilde{\iota}^{(i)}$: the implicit (level-set) function of the bar $\iota^{(i)}$; \\
$\tilde{\omega}$: the implicit (level-set) function of the lattice $\omega$;\\
$\tilde{\Psi}$: the implicit (level-set) function of the lattice $\Psi$.

The bar diameters and endpoints are set as control parameters for lattice unit description and denoted
\eb
\begin{aligned}
\bv &= (\bv^{(1)}_1,\bv^{(1)}_2,\bv^{(2)}_1,\bv^{(2)}_2,\ldots,\bv^{(m)}_1,\bv^{(m)}_2),\\  \p &= (p^{(1)},p^{(2)},\ldots,p^{(n)}).
\end{aligned}
\ee

Each bar $\iota^{(i)}$ consists of a cylinder with diameter $p^{(i)}$ and height $\|\bv_2^{(i)} - \bv_1^{(i)}\|$ and two half-sphere ends with diameter $p^{(i)}$; see Fig.~\ref{pics:single_bar}. The implicit form $\tilde{\iota}^{(i)}$ of bar $\iota^{(i)}$ is defined as follows,
\eb\label{eq-levelset-singlebar}
\tilde{\iota}^{(i)}(\x) =\tilde{\iota}(\x,\bv^{(i)}_1,\bv^{(i)}_2,p^{(i)}) = d(\x,\bv^{(i)}_1,\bv^{(i)}_2) - \frac{p^{(i)}}{2},
\ee
where $d(\x,\bv_1,\bv_2)$ represents the minimum distance from the point $\x$ to the medial axis of the bar,
\eb
d(\x,\bv_1,\bv_2) =
\left\{
\begin{array}{ll}
\|\b\|, &\mbox{if}\ \a\cdot\b\leq 0, \\
\|\g\|, &\mbox{if}\ 0<\a\cdot\b<\a\cdot\a,\\
\|\e\|, &\mbox{if}\ \a\cdot\b\geq \a\cdot\a,
\end{array}
\right.
\ee
with
\eb
\begin{array}{ll}
\a &= \bv_2 - \bv_1\\
\b &=\bv - \bv_1, \\
\e &=\bv - \bv_2,\\
\g &=(\I - \frac{1}{\|\a\|^2}\a\otimes\a)\b.\\
\end{array}
\ee

Accordingly, the implicit function $\tilde{\omega}$ of a lattice unit $\Omega$ is aggregated by the implicit functions $\tilde{\iota}^{(i)}$ ($i=1,2,\ldots,n$) of the $m$ bars. Taking the union of the domains of the bars, that is $\Omega = \cup_{i=1}^m \iota^{(i)}$, we have
\eb
\tilde{\omega}(\x,\bv,\p) = \max_i \tilde{\iota}(\x,\bv_1^{(i)},\bv_2^{(i)},p^{(i)}).
\ee

\begin{figure*}[tb]
\centering
\subfigure[max]{\includegraphics[width=0.15\linewidth]{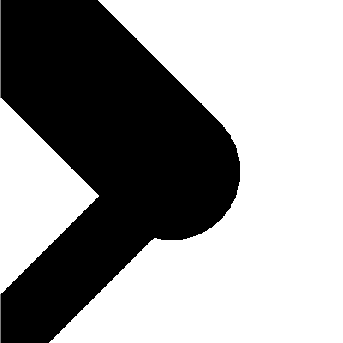}}\quad
\subfigure[KS]{\includegraphics[width=0.15\linewidth]{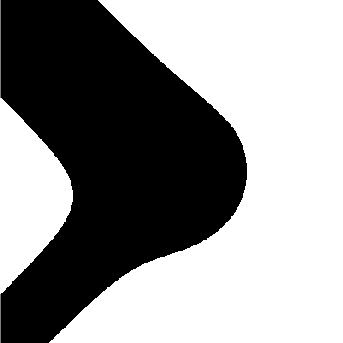}}\quad
\subfigure[difference]{\includegraphics[width=0.15\linewidth]{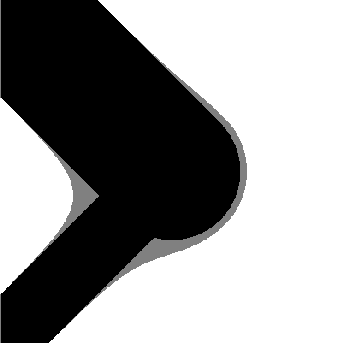}}\quad
\subfigure[max]{\includegraphics[width=0.15\linewidth]{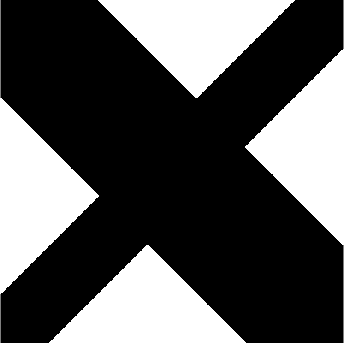}}\quad
\subfigure[KS]{\includegraphics[width=0.15\linewidth]{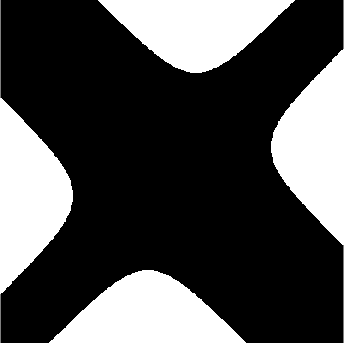}}\quad
\subfigure[difference]{\includegraphics[width=0.15\linewidth]{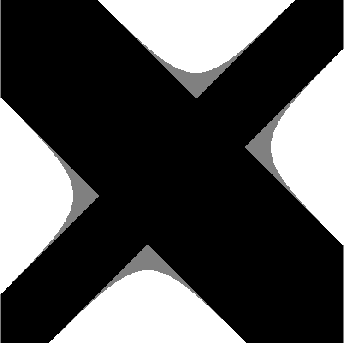}}
\caption{Blending between bars of different widths.}
\label{pics:blendingExamples}
\end{figure*}

To improve the bulk modulus, to reduce stress concentrations and to resolve the problem that the maximum function is not differentiable, the maximum function is replaced by the Kreisselmeier-Steinhauser (KS) function, that is,
\eb\label{eq-KS-blending}
\tilde{\omega}(\x,\bv,\p) = \frac{1}{k}\ln(\sum_i^n e^{k\cdot
\tilde{\iota}(\x,\bv_1^{(i)},\bv_2^{(i)},p^{(i)})}),
\ee
where $k$ is an aggregation coefficient or blending parameter. Fig.\ref{pics:blendingExamples} illustrates the difference between the resulted structures using the maximum function and using the KS function.

\begin{figure}
  \centering\footnotesize
 \subfigure[A lattice unit]{\includegraphics[width=0.18\textwidth]{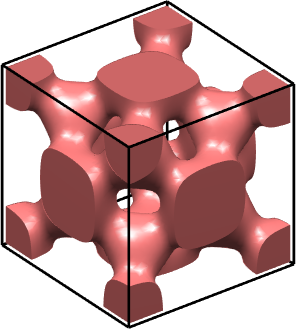}}\quad
 \subfigure[Voxel representation]{\includegraphics[width=0.18\textwidth]{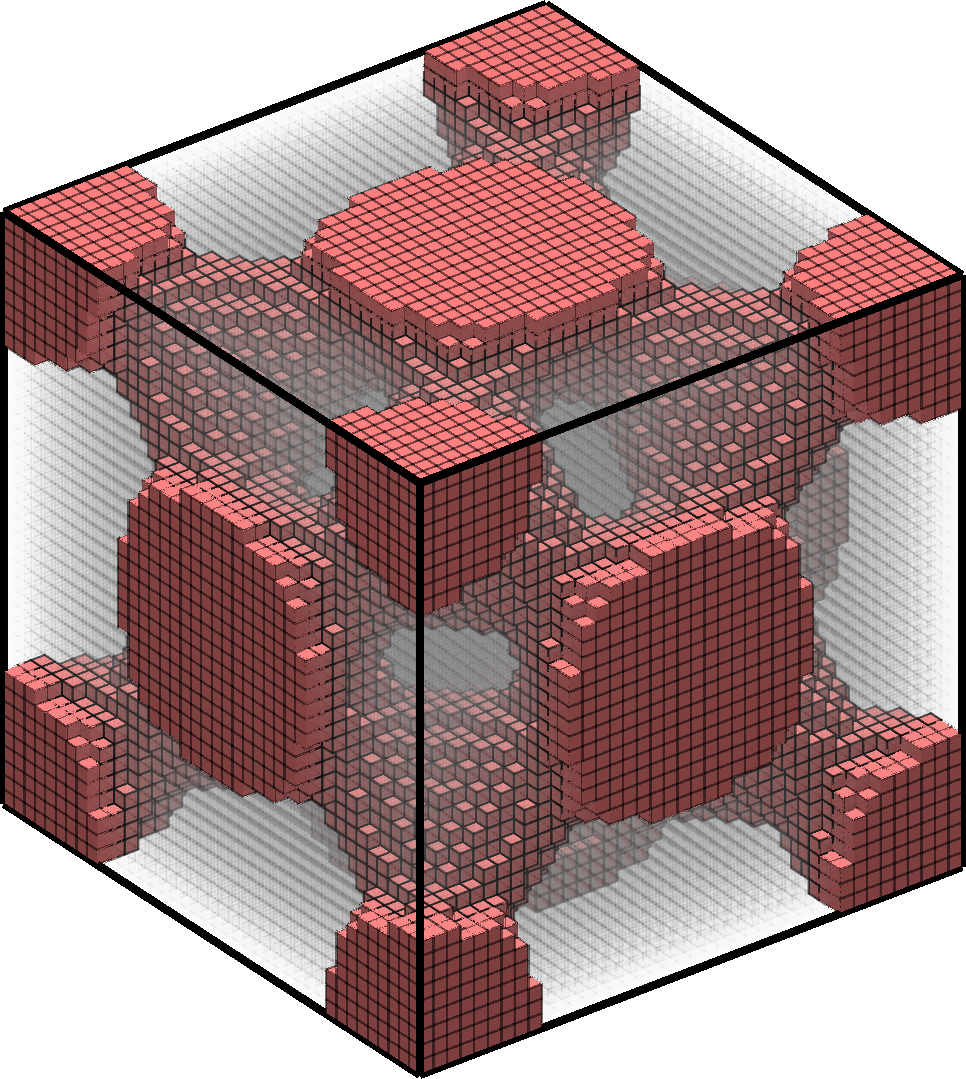}}
  \caption{A lattice unit and its discrete voxel representation.}
  \label{fig-inputs}
\end{figure}

For its FE analysis and a uniform expression of the various types of lattice structures, the design geometry is usually projected onto a fixed regular mesh, being described by a density field as
\eb\label{eq:rho}
\brho(\x) = H(\tilde{\omega}(\x,\bv,\p)) \in [0,1],
\ee
where $H$ is the heaviside function in a regularized version as
\eb\label{eq-H}
H(\tilde{\omega}) =
\left\{
\begin{array}{ll}
0, &\Psi>\gamma, \\
-\frac{3}{4}(\frac{\tilde{\omega}}{\gamma}-\frac{\tilde{\omega}^3}{3\gamma^3})+\frac{1}{2}, &-\gamma \leq \tilde{\omega} \leq \gamma,\\
1,&\Psi<-\gamma,
\end{array}
\right.
\ee
where $\gamma$ is a small positive value controlling the magnitude, and we set $\gamma = 0.005$ in this study. See also Fig.~\ref{fig-inputs}.

Given a solid model $\Omega$ and its hexahedron mesh $\calM$, a lattice structure can then be generated by embedding different units $(\bv,\bp)$ within each mesh element of $\calM$. We also specify $\bP=\{\bP_m\}=\{(\bv_m,\bp_m)\}$ as the control parameters of the lattice structure $\Psi$.

Following a similar procedure as above, the implicit form $\tilde{\Psi}(\bP)$ of the lattice structure $\Psi(\bP)$ can also be defined. Using the implicit modeling, the interior and exterior of the lattice units can be easily determined. It avoids the complex process of meshing and remeshing for FE analysis involved in the optimization iteration process.

\subsection{Basic inverse homogenization}
Given a desired elasticity tensor to a macro-element, we need to find a solid-void microstructure with the desired effective elasticity tensor.
The solution is derived based on the numerical homogenization approach that predicts the material elasticity tensor to a homogeneous microstructure, assuming its periodic distribution
and relatively much smaller size than the macrostructure.

Following the asymptotic homogenization~\cite{andreassen2014determine}, when only the first order terms of the asymptotic expansion
are considered, the homogenized stiffness tensor $D_{ijkl}^H$ is given by averaging the integral over the base cell $Y$ as
\eb\label{eq-homo-1}
D_{ijkl}^{H} = \frac{1}{|Y|} \int_Y D_{pqrs}(\varepsilon_{pq}^{0(ij)}-\varepsilon_{pq}^{*(ij)}) (\varepsilon_{rs}^{0(kl)}-\varepsilon_{rs}^{*(kl)}) {\rm d}Y
\ee
where the Einstein index summation notation is used and $\varepsilon_{pq}^{*(kl)}$ is the $Y$-periodic solution of
\eb
\int_Y D_{ijkl} \varepsilon_{pq}^{*(kl)} \frac{\partial v_i}{\partial y_i} {\rm d}Y
=
\int_Y D_{ijkl} \varepsilon_{pq}^{0(kl)} \frac{\partial v_i}{\partial y_i} {\rm d}Y ,
\ee
where $v$ is $Y$-periodic admissible displacement field and $\varepsilon_{pq}^{0(kl)}$ corresponds to the three (2D) or six (3D) linearly independent unit test strain fields.

Here $ \varepsilon_{pq}^{0(kl)}$ imposes unit test strains (two tensile tests and one shear test) on the boundaries of the base cell, inducing $\varepsilon_{pq}^{A(ij)}$ corresponding to the superimposed strain fields $(\varepsilon_{pq}^{0(ij)}-\varepsilon_{pq}^{*(ij)})$.
Then  Eq.~(\ref{eq-homo-1}) is rewritten in an equivalent
form in terms of element mutual energies
\eb\label{eq-homo-2}
D_{ijkl}^{H} = \frac{1}{|Y|} \int_Y D_{pqrs}(\varepsilon_{pq}^{A(ij)}\varepsilon_{rs}^{A(kl)}) dY.
\ee
In finite element analysis, the base cell is discretized into N
finite elements and Eq.~(\ref{eq-homo-2}) is approximated by
\eb\label{eq-homo-3}
D_{ijkl}^{H} = \frac{1}{|Y|} \sum_{e=1}^{N_e} (\bchi_e^{A(ij)})^T \bk_e \bchi_e^{A(kl)}
\ee
where $\bchi_e^{A(kl)}$ are the element displacement solutions corresponding
to the unit test strain fields $\varepsilon^{0(kl)}$, and $\bk_e$ is the
element stiffness matrix.

The basic inverse homogenization problem is formulated as
\eb
\begin{array}{lll}
\min\limits_{\p}  & \J = || \D^H(\p) - \D^0 ||_F , &  s.t. \\
 & \K^m(\p)\bchi_e^{A(ij)} = \bff^{A(ij)} , & k,l =1, \cdots,d \\
 & V(\p) \leq V^*  & \\
 & \p \in \mathbb{P} \\
\end{array}
\label{eq-problem-inverse_homogenization}
\ee
Here, $\D^0$ is the desired elasticity tensor, $\p$ is the design variable in the admissible space $\mathbb{P}$, $\D^H(\p)$ is the homogenized elasticity tensor defined by Eq.~\eqref{eq-homo-3} and $\K^m(\p)$ is the global stiffness matrix corresponding to design $\p$,
$\bchi_e^{A(ij)}$ is the vector of the nodal displacements to compute and $\bff^{A(ij)}$ the external loadings of the test cases $(kl)$, $V^*$ the desired volume fraction, $V(\brho)$ the volume of the computed structure.

The inverse homogenization Eq.~\eqref{eq-problem-inverse_homogenization} is solved on every group of clustered elasticity tensors to generate the desired microstructure. As the scale separation of the homogenization and the individual optimization of each group, two adjacent different types of microstructures are not necessarily smoothly connected. This is to be improved via a novel physics-based approach, as next explained.

\subsection{Buckling analysis}
With the classical topology optimization formulations for minimum compliance,  designs obtained may contain slender components or high-level stress regions, inducing serious safety risks like structural buckling and material failure.
wich requiring local buckling analysis at the microscale, i.e. the problem of finding
the first local and Y-periodic eigenmode at the length scale of the base cell (highly localized
buckling mode).

The local critical load at the cell level and the respective Y-periodic eigenmodes are obtained by solving the following eigenvalue problem
\eb
\left[
\K^m(\p) - P \G(\bchi, \p)
\right]
\bphi = 0
\ee
$\G$ is the global geometric stiffness matrix.
Specifically,
\eb
\G = - \sum_{e=1}^{N} \int_{Y_e} \frac{\partial \N^\rmT}{\partial y_i} (\sigma_{ij})_e \frac{\partial \N^\rmT}{\partial y_i}  dY
\ee
where $\N$ is the finite element shape function.
Assuming a distinct eigenvalue, the eigenvector is normalized as $(\bchi)^\rmT \K(\p) \bchi = 1$.

The elemental stress $\sigma$ is calculated by
\eb
\bsigma_e = \D_e (\I - \B_e \bchi_e) \bar{\bvarepsilon}
\ee
under the prescribed local strain load $ \bar{\sigma}$ and target elasticity tensor $\bar{E}$
\eb
\bar{\bvarepsilon} = \bar{\D}^{-1} \bar{\sigma},
\ee

where $\bar{\sigma}$  is computed from the macroscale structure consisting of the clustered materials.

Specifically, after clustering the macroscopic problems, each class of materials will correspond to different stresses in different locations. Here, the unit with the largest von Mises stress value in one class is selected and it's element strain is set to the local strain load $\bar{\sigma}$.

\subsection{Inverse homogenization under buckling constraints}
As mentioned before, structural stability and strength requirements should be considered in the optimization. The topology optimization formulation of a compliance minimization problem considering material volume, stress,
and buckling constraints can be stated as
~\cite{Gao2020Improving,Sigmund2018Buckling,neves2002topology}.

The topology optimization of micro inverse homogenization under additional buckling constrains is mainly studied, which is formulated as follows
\eb
\begin{array}{lll}
\min\limits_{\p}  & \J = || \D^H(\p) - \D^0 ||_F , &  s.t. \\
 & \K^m(\p)\bchi_e^{A(ij)} = \bff^{A(ij)} , & k,l =1, \cdots,d \\
 & V(\p) \leq V^*  & \\
 & P_j \geq \underline{P}, &  j \in J_b \\
 & \p \in \mathbb{P} \\
\end{array}
\label{eq-problem-inverse_homogenization_bulkling}
\ee
where

\eb
\left[
\K^m(\p) - P \G(\bchi, \p)
\right]
\bphi = 0
\ee
where $P_j$ is the $j$th buckling load factor corresponding to the given load case, and $\bphi_j$ is the associated buckling mode vector; $J_b$ is a set of indices of the buckling modes considered in the optimization; $\underline{P}$ is the lower bound of buckling load factors.
Specifically, there is
\eb
\G = - \sum_{e=1}^{N} \int_{Y_e} \frac{\partial \N^\rmT}{\partial y_i} (\sigma_{ij})_e \frac{\partial \N^\rmT}{\partial y_i}  dY
\ee

With the original formulation of stress-constrained topology optimization, a large number of stress constraints must be handled due to the local nature of structural stresses. This would make the solution process computationally very expensive in terms of both time and memory. To tackle this difficulty, some single aggregated stress functions, such as the $p$-norm and KS functions, can be used to approximate the maximum stress. While this approach can tremendously reduce the computational cost, it may also introduce approximation errors. As a result, it cannot control the maximum stress accurately. In addition, it increases the nonlinearity of the optimization problem and may cause other convergence difficulties. In the present paper, the KS function and the STM-based correction
scheme are employed~\cite{Gao2020Improving}, and the related formulations are summarized as follows.

For topology optimization of continuum structures considering eigenvalues, mode switching and multimode can appear during the optimization process, which may cause difficulties in sensitivity analysis of eigenvalues. In particular, the repeated eigenvalues associated with multimode are non-differentiable. To circumvent this difficulty, the following single constraint equation is used to replace the original constraint in Eq.\eqref{eq-problem-inverse_homogenization}
\eb\label{eq-fP}
f_P = \underline{P} \kappa^{KS} - 1 \leq 0
\ee
where $\kappa^{KS}$ is a K-S aggregation function defined below
\eb
\kappa^{KS} = \kappa_1 + \frac{1}{\mu_\kappa} \ln
\left[
\sum_{j \in J^\ast_b} \textrm{exp}(\mu_\kappa (\kappa_j - \kappa_1) )
\right]
\ee
in which $\mu_\kappa$ is an aggregation parameter set to $\mu_\kappa = 100/ \kappa_1$; and $J^\ast_b$ is a subset of $J_b$ and contains only indices of the first $n_b$ buckling modes.
To remedy this, a parameter
\eb
c_b =\frac{ \kappa_1}{\kappa^{KS}}
\ee
can be used to scale the aggregation function.

In addition, for the low-density region during optimization, the problem of pseudo buckling modes (pseudo buckling modes) may occur, where we use the method of Gao~\cite{gao2017adaptive}.

Combining the above discussion, the microscopic problem under buckling constraints in this work is re-expressed as
\eb
\begin{array}{lll}
\min\limits_{\p}  & (1-\lambda^B) || \D^H(\p) - \D^0 ||_F  + \lambda^B \kappa^{KS}, &  s.t. \\
 & \K^m(\p)\bchi_e^{A(ij)} = \bff^{A(ij)} , & k,l =1, \cdots,d \\
 & V(\p) \leq V^*  & \\
 & \p \in \mathbb{P} \\
 & \left[ \K^m(\p) - P \G(\bchi, \p) \right] \bphi = 0
\end{array}
\label{eq-problem-inverse_homogenization_bulkling}
\ee
where $\lambda^B$ is the weight of the buckling constraint

\subsection{Numerics and sensitivities}

The lattice microstructure is projected onto density field for FE analysis as described in Sec.~\ref{sec-latmodel}.
And the constitutive property $E$ is interpolated between a void and solid phase using the modified SIMP scheme as

\eb
E(\rho) =
\left\{
\begin{array}{ll}
  E_{\min} + \rho^p (E_0 - E_{\min}) & for\ \K^m,\f,  \\
   \rho^p E_0, & for\ \G,
\end{array}
\right.
\label{eq:dp}
\ee
where $p$ is a penalization factor, $E_{\min}$ is the relative density of the void phase which is set to a small value to avoid numerical singularities, and $E_0$ is the constitutive tensor of the solid phase.

The modified scheme is used to eliminate the notoriously known problem of artificial modes which contaminate low density regions. To ensure that the scheme is successful, the value of $E_{\min}$ must be sufficiently large to stabilize low density regions while still being small enough to have negligible influence on the bifurcation loads. Through multiple numerical studies, $E_{\min} = 10^{-4} E_0$ was found to yield the best results without causing significant alterations of the loads.

In the specification of the design domain, symmetry constraints in the material distribution are utilized. These geometric restrictions will enforce material symmetries in the material properties (e.g. square symmetry or isotropy), but also substantially lower the computational cost of the stability analysis as it reduces the unique part of the Brillouin zone. Accordingly, element densities within the fundamental domain will be designated as design variables $\mu$, and element densities of the complete domain $\rho_{\min}$ are obtained through the mapping:
\eb
\rho = \L \mu,
\ee
where $\L$ is a mapping matrix with dimension $N \times n$. 

Computing solution to the inverse homogenization problem~\eqref{eq-problem-inverse_homogenization_bulkling} is very challenging due to the inclusion of the nonlinear objective and constraint functions, which results in an overall complex nonlinear and nonconvex optimization problem. Due to this considerations, the well-established optimization approach Globally Convergent Method of Moving Asymptotes (GCMMA)~\cite{zillober1993Globally} is applied here. It approximates the original nonconvex problem through a set of convex subproblems by using the gradients of the optimization objective and constraints with respect to the design variables.

The approach mainly depends on computation of the derivatives of the objective and constraint functions. Different from the density-based topology optimization, the design variables of the microstructure optimization are the widths of the bars instead of the discretized density field.
The sensitivities of the objective and constraint functions with respect to the design variables are computed via the chain rule as
\eb
\frac{d~f(\p)}{d~\p} = \frac{\partial~f(\p)}{\partial~\rho} \frac{\partial~\rho}{\partial~\p},
\ee
where $f(\p)$ represents the objective function or the constraint function, i.e. $f(\p) = \bJ, V(\p)$ defined in Eq.~\eqref{eq-problem-inverse_homogenization_bulkling} or $f_P$ defined in Eq.~\eqref{eq-fP}. The gradients of the density filed with respect to the design variables ${\partial~\rho}/{\partial~\p}$ can be derived from Eqs.~\eqref{eq-H}, \eqref{eq-levelset-singlebar} and \eqref{eq-KS-blending}

and the sensitivities with respect to the density field ${\partial \bJ}/{\partial \rho_e}$, ${\partial V}/{\partial \rho_e}$ and ${\partial f_P}/{\partial \rho_e}$ are derived as follows.

The derivatives of the objective function $\bJ$ with respect to density $\rho_e$ is derived as,
{\small
\begin{align}
&\frac{\partial \bJ}{\partial \rho_e} = \frac{\partial \bJ}{\partial \bD^H}\frac{\partial \bD^H}{\partial \rho_e},
\end{align}}
where

{\small
\begin{align}
&\begin{aligned}
\frac{\partial \bJ}{\partial \bD^H} = \Tr((\bD^H(\brho)-\bD^0)^T&(\bD^H(\brho)-\bD^0))^{(-\frac{1}{2})}\cdot\\
                                                                &(\bD^H(\brho)-\bD^0),\\
\end{aligned}\\
&\frac{\partial \bD^H}{\partial \rho_{e}} = \frac{1}{|\Omega_e|} p \rho_{n}^{p-1}\left(E_{0}-E_{\min }\right)\left(\bmu_{n}^{A(ij)}\right)^{T} \bK_0 \bmu_{e}^{A(kl)}.
\end{align}}

The derivatives of the volume constraint $V$ with respect to density $\rho_e$ is derived as
\eb
\frac{\partial V(\brho)}{\partial \rho_{e}} = 1.
\ee

And the sensitivity of the buckling constraint function with respect to $\rho_e$
can be expressed as
\eb
\frac{\partial f_P}{\partial \rho_e} = \underline{P} \frac{\partial \kappa^{KS}}{\partial \rho_e} = \frac{\underline{P}}{\sum_{j \in J^\ast_b } \textrm{exp}(\mu_\kappa \kappa_j) }
{\sum_{j \in J^\ast_b } \textrm{exp}(\mu_\kappa \kappa_j) \frac{\partial \kappa_j}{\partial \rho_e} }.
\ee
If the eigenvalue $\kappa_j$ is unimodal, its sensitivity with respect to $\rho_e$ reads
\eb
\frac{\partial \kappa_j}{\partial \rho_e} =
\bphi^\rmT_j \left(
\frac{\partial \G}{\partial \rho_e}
-
\kappa_j \frac{\partial \K^m}{\partial \rho_e}
\right) \bphi_j
-
\sum_{k=1}^{3}  {\bv_j^k}^\rmT
\left( \frac{\partial \K^m}{\partial \rho_e} \bchi^k - \frac{\partial \f^k}{\partial \rho_e} \right)
\ee
where $\bv_j^k$ is an adjoint vector with respect to the $j$th buckling mode.
In 2D cases, we note that $11 \rightarrow 1$,
$22 \rightarrow 2$, and $12 \rightarrow 3$,
For load $k=1,2,3$ and buckling modes $\bphi_j ( j = 1,..,n_b)$, the above vectors can be obtained by solving the following adjoint equation
\eb
\K^m \v_j^k = \P_j^k
\ee
with the right-hand-side vector $\P_j^k$ given as

\eb
\begin{array}{rcl}
  \P_j^k & = & {\bphi_j^k}^\rmT \frac{\partial \G}{\partial \bchi^k} {\bphi_j^k}  \\
   & = & \sum_{e=1}^{N}\left\{
   {\bphi_{j,e}^k}^\rmT  \frac{\partial \G_e}{\partial \bchi^k_{e,1}}  {\bphi_{j,e}^k}, \cdots,
   {\bphi_{j,e}^k}^\rmT  \frac{\partial \G_e}{\partial \bchi^k_{e,8}}  {\bphi_{j,e}^k}
    \right\}^\rmT
\end{array}
\ee
where
\eb
\frac{\partial \G_e}{\partial \chi^k_{e,i}} =
- \int_{Y_e} \frac{\partial N^\rmT}{\partial y_l}
\frac{\partial (\sigma_{lm})_e }{\partial \chi^k_{e,i}}
\frac{\partial N^\rmT}{\partial y_m}  dY
\ee
and
\eb
\frac{\partial \bsigma_e }{\partial \chi^k_{e,i}} = - \E_e \B_e \e_{ik} \bar{\bvarepsilon}, \qquad i=1,\cdots,8.
\ee

The derivatives of the global stiffness matrices $\K^m$, $\G$ and loads $\bf_j$  are calculated by
  \eb
  \frac{\partial \G}{\partial \rho_e} = p \rho^{p-1}_e \G_e^0
  \ee

  \eb
  \frac{\partial \K^m}{\partial \rho_e} = p (1- E_{\min}/E_0) \rho_e^{p-1} \bk_e^0.
  \ee

  \eb
  \frac{\partial \f^k}{\partial \rho_e} = p (1- E_{\min}/E_0) \rho_e^{p-1} \f_{e}^{0k}.
  \ee
where $\bk_e^0$, $\G_e^0$, $\f_{e}^{0k}$ are element matrices and vectors evaluated with the constitutive matrix of the solid phase $\E_0$.

\eb
\frac{\partial~\rho}{\partial~ p^{(i)}} = \frac{\partial~H(\Psi(\p))}{\partial~ p^{(i)}}
= \frac{\partial~H}{\partial~\Psi} \frac{\partial~ \Psi(\p)}{\partial~ p^{(i)}}
\ee
\eb
\frac{\partial~H}{\partial~\Psi} =
\left\{
\begin{array}{ll}
\frac{3(\varepsilon-1)(\Psi^2-\gamma^2)}{4\gamma^3}, & -\gamma \leq \Psi \leq \gamma,\\
0,& \Psi<-\gamma\ \mbox{or}\ \Psi>\gamma,
\end{array}
\right.
\ee
\eb
\frac{\partial~ \Psi(\p)}{\partial~ p^{(i)}}
=-\frac{\exp(-k(d^{(i)}-\frac{p^{(i)}}{2}))}{\sum_{j=1}^{n} \exp(-k(d^{(j)}-\frac{p^{(j)}}{2}))}
\ee

\subsection{Post-processing}\label{sec-bucktheory-connect}

The microstructure obtained by solving the aforementioned optimization problem may have some structures that do not meet the actual manufacturing requirements, for example, the rods in a single microstructure are too thin and overhanging, and there are also cases where adjacent units are not connected when the microstructure is stitched together as a whole.
To address the problem, this subsection proposes the following steps to post-process the microstructure.

1. Structural rationality check of the over-thin rods, which need to be removed due to the cost of actual manufacturing and process considerations. This operation may affect the connectivity of the structure inside the unit, so the structural connectivity must be checked and the excess suspension rods must be removed. Here, the connectivity of the discrete unit structure is determined based on eight units.

2. The above-mentioned rod volume optimization: based on the volume constraint, the radius of all existing rods is optimized twice, so that the total volume of the structure meets the constraint requirements.

3. Connectivity of adjacent units: global structure splicing, connectivity detection and optimization of adjacent units. For two adjacent units that are not connected, for the two rods with the closest endpoint distance, a rod is added in the middle to connect the end of the original two rods, and the radius of the new rod is smoothly transitioned between the two by space.


\section{Examples}\label{sec-example}

The proposed method has been implemented in MATLAB with an experimental computer environment of Intel Core i5-4590 3.2GHz CPU and 16GB RAM. the FMO problem is solved by YALMIP~\cite{lofberg2004yalmip} calling MOSEK 9.1. The linear objective functions and constraints other than the buckling problem form a good mathematical structure, but introduce additional large-scale matrix inequalities whose computation is very time-consuming and requires large memory resources. A more efficient approach ~\cite{weldeyesus2015primal} can be used to solve specific FMO problems. In all examples, the inverse homogenization problem ~\eqref{eq-problem-inverse_homogenization_bulkling} has a design region size of $50\times50$, a small cell length of $1\times1$, and a microstructure composed of materials with Poisson's ratio $\nu=0.3$, Young's modulus $E_{0} = 1$, $E_{min} = 0.1^3$ . The macroscopic and each microscopic volume constraint is uniformly set to 0.35 in this section.

\begin{figure}[t]
\centering
{\includegraphics[width=0.45\textwidth]{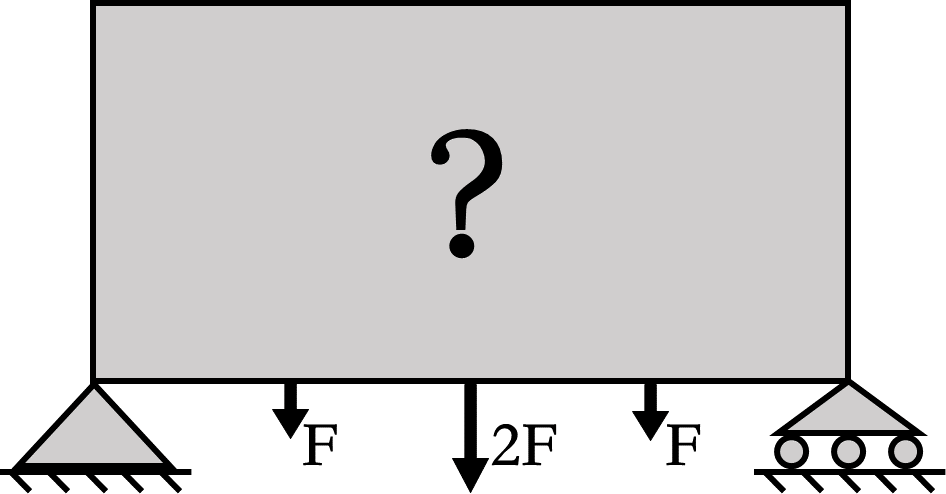}}
\caption{Bridge.}
\label{buckpic-problem}
\end{figure}

In this chapter, the performance of the proposed method is tested in terms of different microscopic lattice structure models and different flexural constraint weights.
First section \ref{bucksec-eMic} gives the optimization results for a single microstructure and shows the effectiveness of the proposed connection strategy for different microstructures in this chapter. Then, Sections \ref{bucksec-halfmbb} and \ref{bucksec-bridge} test the optimization of the bridge problem shown in ~\ref{buckpic-problem}, respectively, in each homogeneous material space, when using different microstructure models and under different weighted buckling constraints.

\subsection{Design of microstructure}\label{bucksec-eMic}

This subsection gives the effect of optimal design for a single microstructure.
The basic model of microstructure shown in Figure ~\ref{buckpic-micModel} is used for the quiz in this chapter.
The figure shows the basic model of the one-eighth lattice microstructure, taking three vertices, five edge points and one face point, and using the 21 rods identified by these nine points as design variables.

\begin{figure}[t]
\centering
{\includegraphics[width=0.45\textwidth]{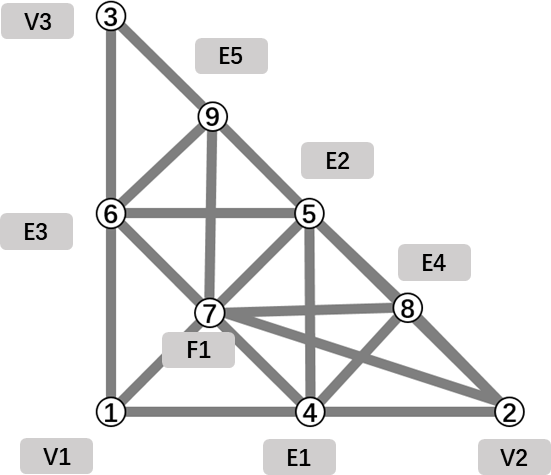}\label{buckpic-micModel}}
\caption{The basic model of the one-eighth lattice microstructure uses 21 rods determined by 9 points as design variables, and the thickness of each rod is determined by optimization.}
\label{buckpic-micModel}
\end{figure}

\begin{figure}[hpt]
\centering
(a)
\begin{minipage}[c]{0.435\textwidth}
{\includegraphics[width=0.335\textwidth]{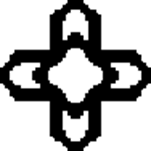}\label{buckpic-hm-wei0-type1}}
{\includegraphics[width=0.535\textwidth]{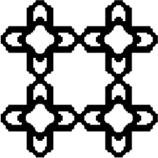}\label{buckpic-hm-wei0-type1-4}}
\end{minipage}
(b)
\begin{minipage}[c]{0.435\textwidth}
{\includegraphics[width=0.335\textwidth]{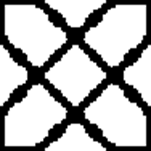}\label{buckpic-hm-wei0-type2}}
{\includegraphics[width=0.535\textwidth]{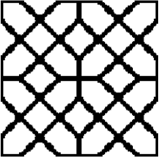}\label{buckpic-hm-wei0-type2-4}}
\end{minipage}
(c)
\begin{minipage}[c]{0.435\textwidth}
{\includegraphics[width=0.335\textwidth]{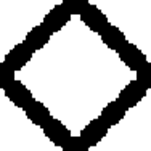}\label{buckpic-hm-wei0.02-type3}}
{\includegraphics[width=0.535\textwidth]{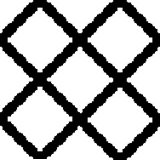}\label{buckpic-hm-wei0.02-type3-4}}
\end{minipage}
(d)
\begin{minipage}[c]{0.435\textwidth}
{\includegraphics[width=0.335\textwidth]{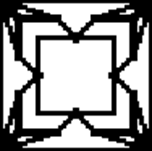}\label{buckpic-hm-wei1-type5}}
{\includegraphics[width=0.535\textwidth]{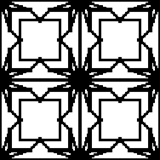}\label{buckpic-hm-wei1-type5-4}}
\end{minipage}
(e)
\begin{minipage}[c]{0.435\textwidth}
{\includegraphics[width=0.335\textwidth]{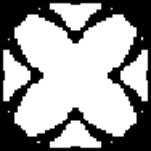}\label{buckpic-bridge-wei0.4-type5}}
{\includegraphics[width=0.535\textwidth]{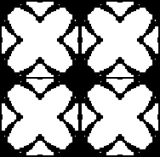}\label{buckpic-bridge-wei0.4-type5-4}}
\end{minipage}
(f)
\begin{minipage}[c]{0.435\textwidth}
{\includegraphics[width=0.335\textwidth]{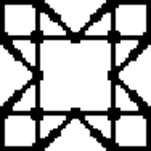}\label{buckpic-bridge-wei1-type4}}
{\includegraphics[width=0.535\textwidth]{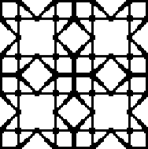}\label{buckpic-bridge-wei1-type4-4}}
\end{minipage}
\caption{Example of microstructure.}
\label{buckpic-micExample}
\end{figure}

\begin{figure}[hpt]
\centering
\subfigure[Case 1 of connection related to structure a.]
{\includegraphics[width=0.3\textwidth]{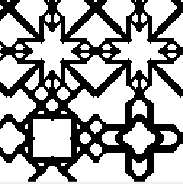}\label{buckpic-micconnect-a1}} \ \ \ \ \ \ \ \
\subfigure[Case 2 of connection related to structure a. ]
{\includegraphics[width=0.3\textwidth]{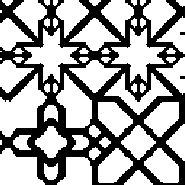}\label{buckpic-micconnect-a2}} \\
\subfigure[Case 1 of connection related to structure b.]
{\includegraphics[width=0.3\textwidth]{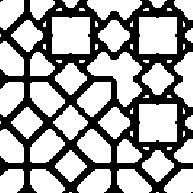}\label{buckpic-micconnect-b1}} \ \ \ \ \ \ \ \
\subfigure[Case 2 of connection related to structure b. ]
{\includegraphics[width=0.3\textwidth]{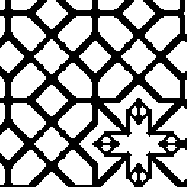}\label{buckpic-micconnect-b2}}
\caption{Example of microstructure connection.}
\label{buckpic-micconnect}
\end{figure}

Figure ~\ref{buckpic-micExample} gives the optimization results of six lattice microstructures, and the graph shows the effect of a single lattice structure and its four-cell tiling at $60 \times 60$ resolution.
In practice, there may be a problem that the adjacent different structural cells cannot be directly stitched together, and the method proposed in section \ref{sec-bucktheory-connect} is used here to deal with it. Figure ~\ref{buckpic-micconnect} gives some connection examples of structure (a) and structure (b) in Figure ~\ref{buckpic-micExample}. It can be seen that the scheme can ensure continuity when splicing different structures.

\subsection{Basic bridge problem}\label{bucksec-bridge}

\begin{figure}[hpt]
\centering
\subfigure[Results of macroscopic optimization in a continuous free isotropic space, $c=1.4618$.]
{\includegraphics[width=0.5\textwidth]{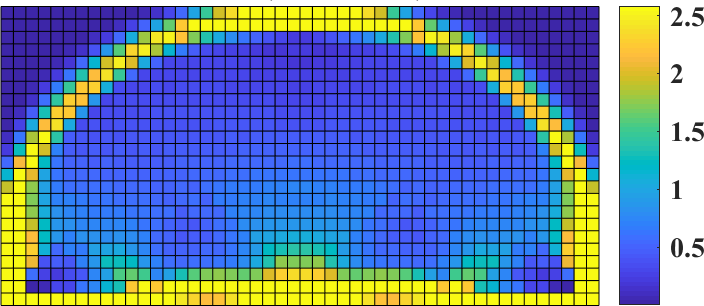}\label{buckpic-bridge_FIMO}}
\subfigure[Results of clustering$c=1.5899$. Number of clusters is 5.]
{\includegraphics[width=0.42\textwidth]{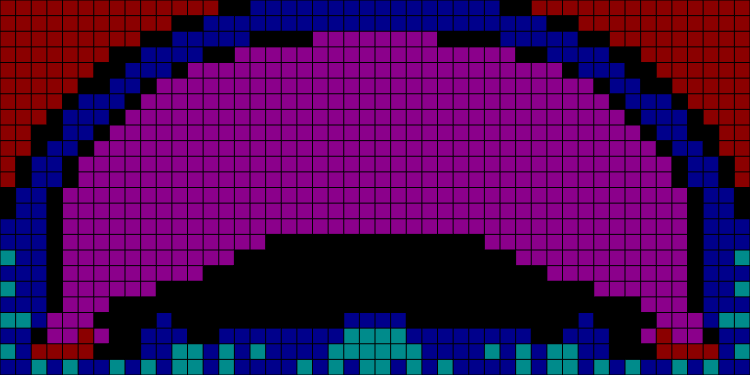}\label{buckpic-bridge_FIMO_cluter}}
\caption{Macro-optimization results for the first step of the bridge problem.}
\label{buckpic-bridge-sol}
\end{figure}

\begin{figure}[hpt]
\centering
\subfigure[$\lambda^B=0$，$c =  19.6117$。]
{\includegraphics[width=0.5\textwidth]{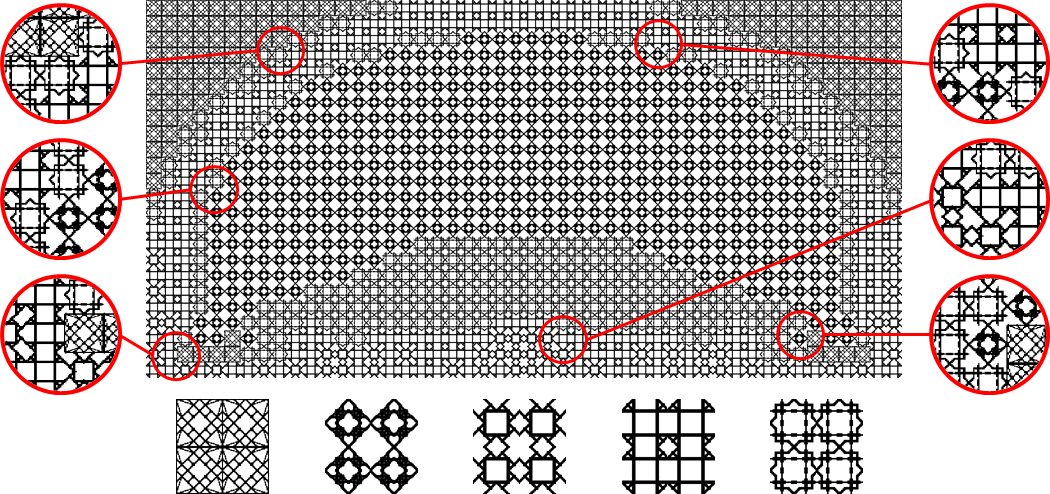}\label{buckpic-bridge-0}}
\subfigure[$\lambda^B=0.01$，$c =  14.0077$。]
{\includegraphics[width=0.5\textwidth]{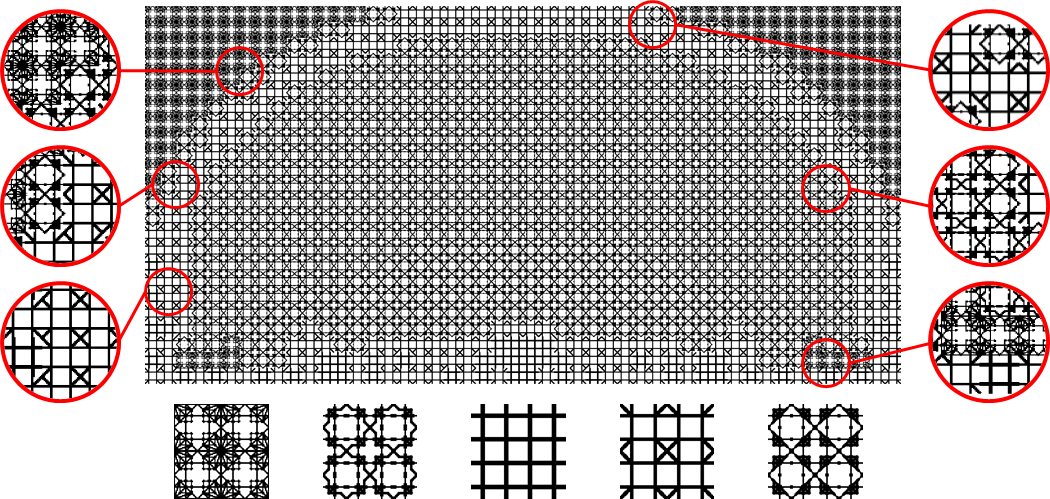}\label{buckpic-bridge-0.01}}
\subfigure[$\lambda^B=0.02$，$c =  23.9039$。]
{\includegraphics[width=0.5\textwidth]{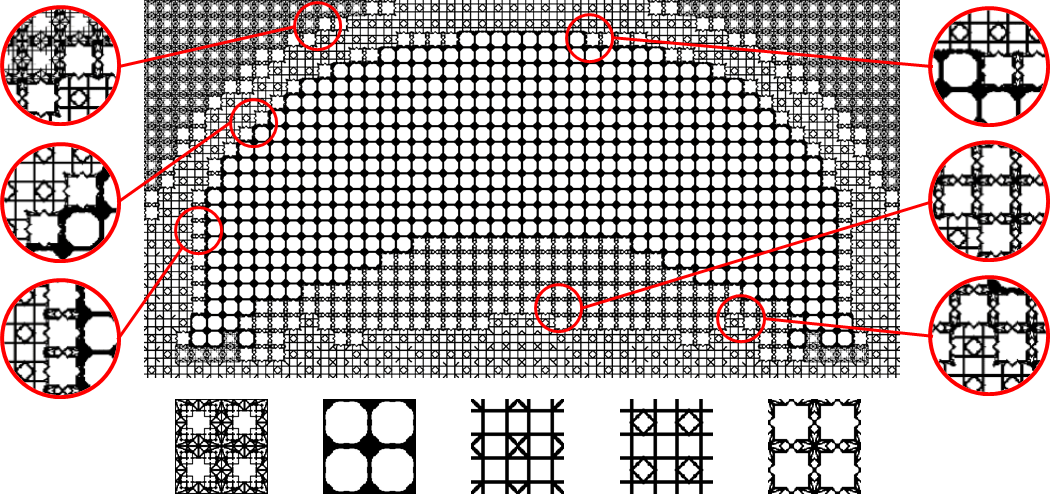}\label{buckpic-bridge-0.02}}
\subfigure[$\lambda^B=0.4$，$c =  24.9216$。]
{\includegraphics[width=0.5\textwidth]{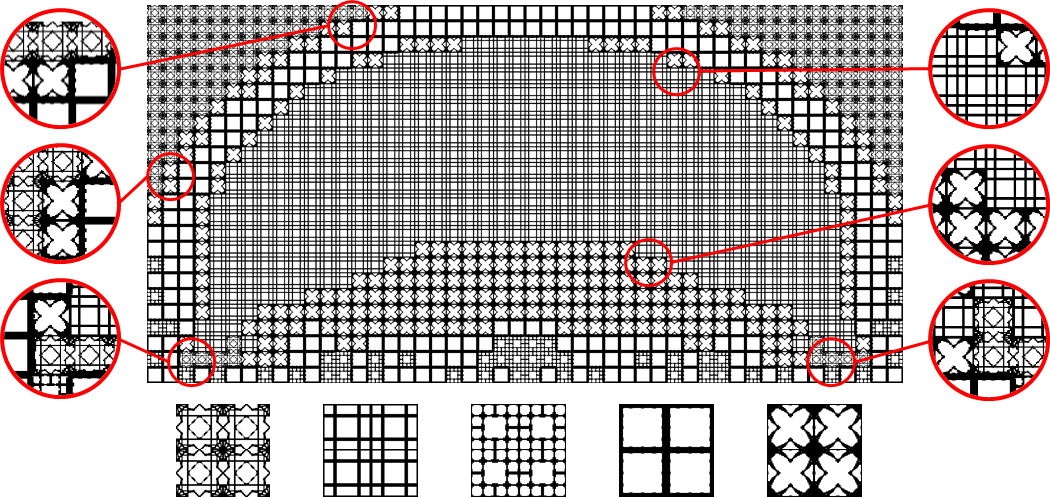}\label{buckpic-bridge-0.4}}
\subfigure[$\lambda^B=0.9$，$c = 11.4800$。]
{\includegraphics[width=0.5\textwidth]{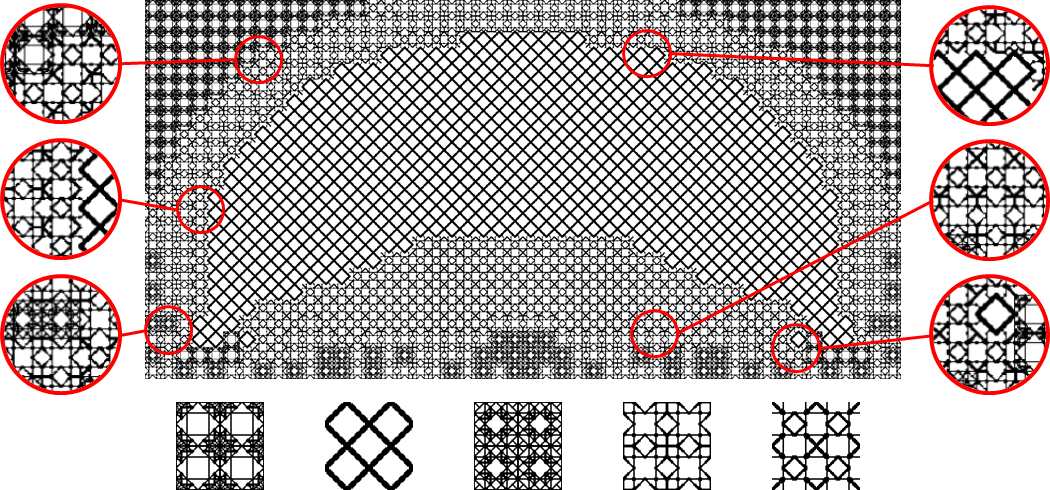}\label{buckpic-bridge-1}}
\caption{Bridge problem.}
\label{buckpic-bridge}
\end{figure}

This subsection gives the optimization results for the classical bridge problem shown in Figure ~\ref{buckpic-problem}.
The macroscopic structure is partitioned into $48 \times 24$ cells, with an external load size of 0.1, a volume constraint of 0.35, and the number of clusters set to 5.
The microstructure optimization resolution is $40 \times 40$, and the microscopic lattice optimization model is adopted from the structure shown in Figure ~\ref{buckpic-micModel} in Section \ref{bucksec-eMic}.
Considering that we are using a symmetric design strategy for the one-eighth lattice, here we optimize the macroscopic problem in isotropic isotropic space. Figure ~\ref{buckpic-bridge-sol}(a) shows the result of macroscopic successive optimization with the flexibility $c=1.4618$, which is clustered to obtain the structure consisting of five materials with the flexibility $c=1.5899$ shown in Figure ~\ref{buckpic-bridge-sol}(b).
Based on this, the microstructures corresponding to the species materials are designed separately and different flexural constraint weights are applied, and the results are shown in Fig. ~\ref{buckpic-bridge}. Here the flexural weights $\lambda^B$ take the values of 0, 0.01, 0.02, 0.4, and 0.9, respectively, in turn.
It can be seen that the overall structural performance does not decrease monotonically due to the introduction of local buckling constraints, which may be caused by the local nature of the inverse homogenization algorithm on the one hand, and the local buckling constraint has a positive contribution to the overall structure, which can overcome the aforementioned localization to a certain extent.

\subsection{self-supporting bridge problem}\label{bucksec-halfmbb}

\begin{figure}[t]
\centering
{\includegraphics[width=0.45\textwidth]{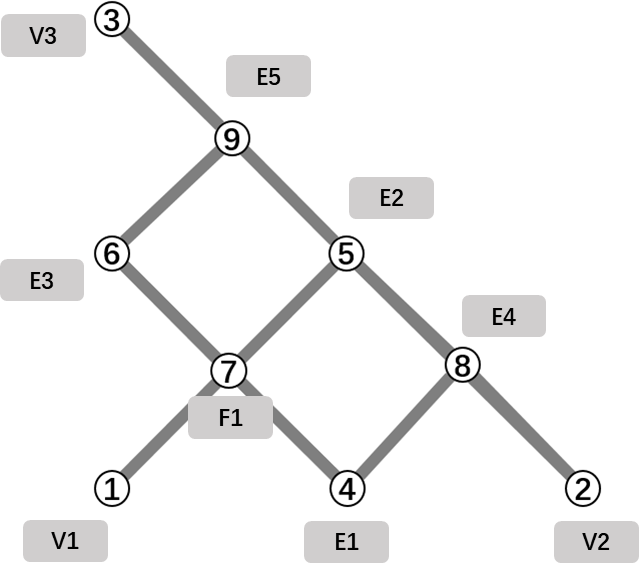}}
\caption{The basic model of the one-eighth lattice for the self-supporting microstructure uses 10 rods determined by 9 points as design variables to optimize the determination of the thickness of each rod.}
\label{buckpic-micModel45}
\end{figure}

\begin{figure}[hpt]
\centering
\subfigure[$\lambda^B=0$，$c = 28.0892$。]
{\includegraphics[width=0.5\textwidth]{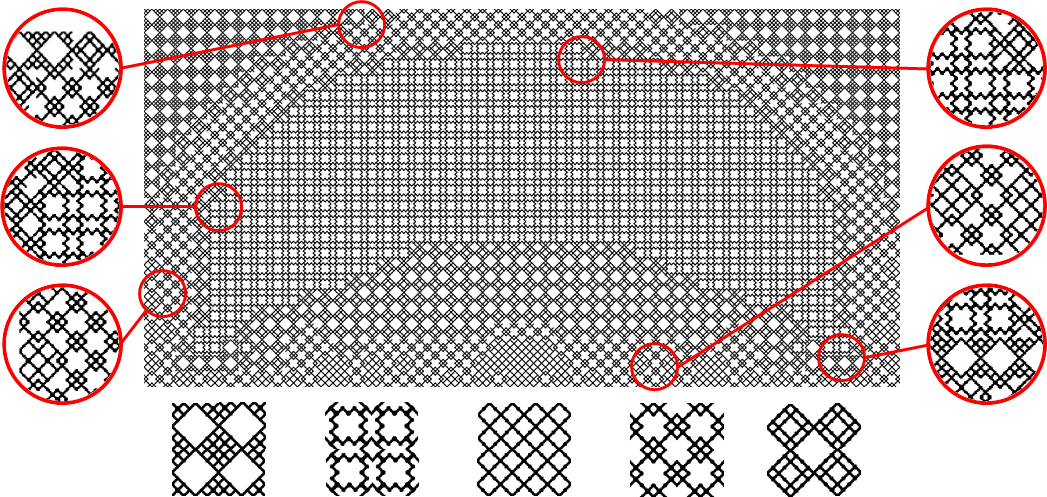}\label{buckpic-bridge-0}}
\subfigure[$\lambda^B=0.01$，$c = 25.2472$。]
{\includegraphics[width=0.5\textwidth]{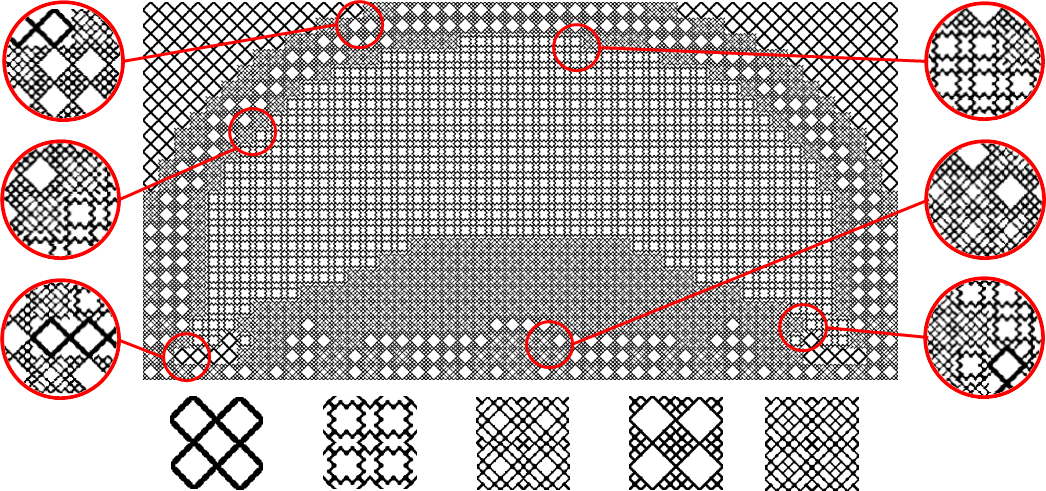}\label{buckpic-bridge-0.01}}
\subfigure[$\lambda^B=0.02$，$c = 28.4533$。]
{\includegraphics[width=0.5\textwidth]{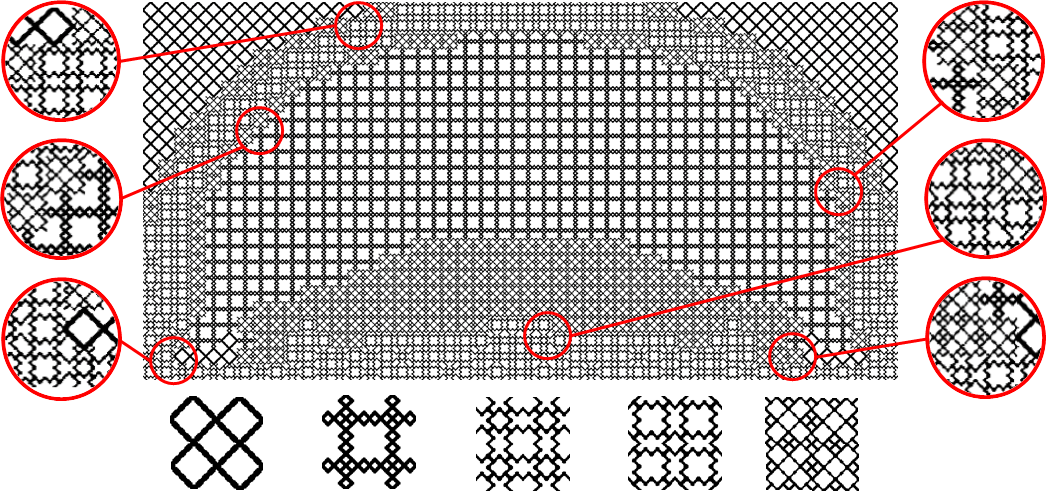}\label{buckpic-bridge-0.02}}
\subfigure[$\lambda^B=0.4$，$c =  22.3849$。]
{\includegraphics[width=0.5\textwidth]{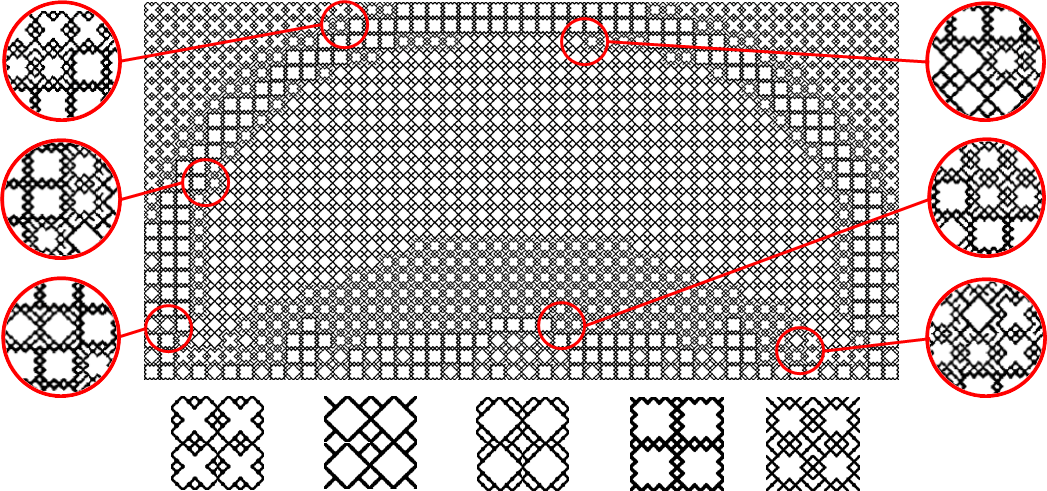}\label{buckpic-bridge-0.4}}
\subfigure[$\lambda^B=0.9$，$c = 25.2772$。]
{\includegraphics[width=0.5\textwidth]{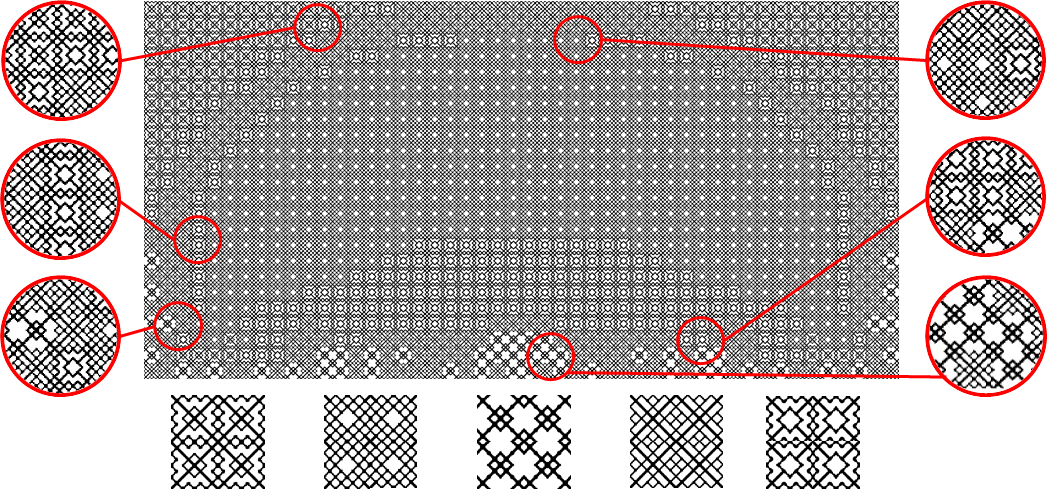}\label{buckpic-bridge-1}}
\caption{Self-supporting optimization results for bridge problem.}
\label{buckpic-bridge45}
\end{figure}

Different from the previous subsection, this section optimizes the bridge problem in the self-supporting microstructure space, so that the resulting structure naturally conforms to the self-supporting requirements of the 3D printer material. The approach used here is to directly introduce the basic model of the self-supporting lattice, as shown in Figure ~\ref{buckpic-micModel45}, where each rod is at 45 degrees to the printing direction, thus making the final obtained overall structure self-supporting.
Figure ~\ref{buckpic-bridge45} shows the optimization results for this problem with different buckling constraint weights, and the overall structural flexibility is 28.0892, 25.2472, 28.4533, 22.3849, and 25.2772, in that order.
Compared with the results of ~\ref{buckpic-bridge} in the previous section, the introduction of the self-supporting constraint causes a significant decrease in the overall flexibility performance of the structure, but it is generally within an acceptable range and does not incur additional computational cost.

\section{Conclusion}
This paper proposes and implements a lattice structure optimization design method that satisfies the local buckling requirements, enabling the selective filling of several different kinds of lattice structures inside the model to satisfy the overall rigidity and local buckling requirements. This method uses a strategy based on free material optimization to obtain the overall optimal material elastic tensor distribution and stress tensor at each cell to the maximum extent possible. Based on this, the inverse homogenization method is further extended to achieve a cell lattice structure that satisfies the flexural constraints as well as the matching elastic tensor. In contrast to the topology-based optimization method, the lattice structure shape parameters are used as design variables to ensure the geometric validity of the final lattice structure and the convergence of the optimization algorithm. The effectiveness of the algorithm is finally verified by various numerical experiments.

\bibliographystyle{elsarticle-num}
\bibliography{BUCKLING_bib,citations_Neves2002Topology,HomCNN}

\end{document}